# Microgrid Reference Methodology for Understanding Utility and Customer Interactions in Microgrid Projects

Franklin E. Pacheco, *Graduate Student Member, IEEE*, J. Chris Foreman, *Senior Member, IEEE*

*Abstract* – The purpose of this paper is to introduce a holistic and systematic approach, based in key concepts of systems thinking, systems of systems and management science; to completely represent, model and analyze microgrid systems. In this paper we propose a Microgrid Reference Methodology (MRM) that guides the cooperation and mutual benefits between electric utilities and industrial customers for microgrid projects. We sketch a four-level socio-technical system in which the δ (business) level is added to the other three levels that are traditionally analyzed in microgrid design and planning. This MRM clearly specifies the steps and processes necessary for determining actors in the system, their interrelations, interests, goals and undesired effects. Finally, the MRM makes possible the evaluation of the impact of different alternatives on the objectives of both parties through the determination of criteria and factors. These factors can be influenced by the electric utility and customer; or they can be external, but still influenced by other actors such as regulators and government, to incentivize the implementation of microgrid projects.

## I. INTRODUCTION

IN recent years, the energy market has experienced important challenges in its structure and in the requirements of its actors such as the necessity for more reliable electric service, energy efficiency, environmental care practices, and the decentralization of power generation incorporating renewable energies and Distributed Energy Resources (DER). Given this context, microgrids perform different functions that offer several benefits to the electrical grid in terms of system robustness, power quality, reliability, and security [1] [2]. As a localized electrical grid with the capability of islanding itself when there is a power quality event in the main grid, a microgrid includes its own generation sources and a management, operation and control system (MOCS) to deliver increased power resilience to critical loads, and manage the intermittent DER.

The adoption of microgrid technology is expected to grow rapidly in the United States of America (USA), as well as in various other countries of the world. In addition to traditional users of microgrids such as governmental agencies and the military, microgrids are becoming attractive to big consumers such as factories, supermarkets, universities and hospitals. The market opportunity for microgrids is expected to grow by more than 3.5 times [3] between 2015 and 2020.

Despite the benefits and positive market projections of microgrids, few actual implementations have been performed until now [4] [5] and there is a lack of participation of electric utilities. There are several concerns that limit their participation in this kind of projects [6]. For instance, some utilities think their corporate profits would be threatened by customer microgrid projects. Utilities cannot see incentives and attractive paybacks easily, and worry about power quality impacts from customer microgrids

There have been great advances in developing cheaper and more efficient technologies to take advantage of the DER, and to optimize the microgrid system. However, most of the past and current research has focused mainly on solving technical challenges and few of them have addressed an additional dimension, the microgrid market, in which the interactions between the actors in a microgrid project are analyzed in order to understand the socio-technical complexity of microgrids and to propose solutions that benefit all the actors.

## II. BACKGROUND

There are several articles that focus on microgrid optimization and decision-making; most are focused on technical aspects and purely electrical variables; however, only a few related with the perspective of a microgrid from a holistic view. For instance, in [7] it is used an iterative Systems Engineering (SE) life-cycle to describe the stakeholder analysis and requirement analysis for the development of rural microgrids. In [8] it is presented a Microgrid Commercial Ecosystem and Analysis Components tool to visualize the main microgrid participants and their interactions. Finally, a model of the economic interactions between microgrids and electric utilities from a Regulator perspective are presented in [9] using a cooperative game theory to determine the revenues and showing when microgrid development might be beneficial.

The Microgrid Reference Methodology (MRM) proposed in this paper presents an actual process and tools to determine the electric utility – customer interactions based on different frameworks and concepts from the fields of Systems Engineering, System of Systems, Management Science, and Infrastructure Architectures. Hence, a brief explanation of those concepts and the usefulness for the analysis of microgrids is discussed here.

### A. Systems Thinking

"Systems thinking is a discipline for seeing wholes ... for seeing interrelationships rather than things, for seeing patterns rather than static snapshots. It is a set of general principles spanning fields as diverse as physical and social sciences, engineering and management" [10], In other words, systems thinking considers the whole rather than the parts in an interdisciplinary approach to solve complex real world problems which do not have one simple answer.

The literature on systems thinking is widely applied in different fields of knowledge to analyze complex problems for which a mathematical equation will not necessarily obtain the best solution. In engineering, SE defines what is to be done by creating requirements, concepts, and architectures that will be used by functional engineering. SE focuses on the architecture and the starting point is determining the user requirements [8].

The systematic approach to system design follows a life cycle. There are different approaches such as the linear, evolutionary, Waterfall, Spiral, and Vee models which include the stages of user requirements, specifications of the system, implementation, integration and testing, operation and deployment [12].

These life cycles and holistic thinking are important to analyze microgrids as systems; however, microgrids are composed by many systems and dimensions, hence the concepts of Systems of Systems (SoS) are relevant.

### B. Systems of Systems

SoS is a relatively new special class of systems. The term first appeared in 1989, but the concept has not been completely clear until recently [13]. After an iterative process and the collaboration of different researchers, some agreement exists on a SoS: It is constituted by components which individually may be considered as systems, the behavior of the SoS is not obtained from any individual component, and the components are operationally and managerially independent. There is no directed or governing structure. Instead, it is a collaborative environment [14], there is a significant complexity and heterogeneity, there is an emergent behavior that cannot be analyzed by dividing the analysis in parts, the interactions and relations between systems are crucial for analysis.

For the MRM, the SoS analysis helps to understand the nature of a microgrid as not just a specific technology, but a collection of different elements from non-centralized electric power sources such as photovoltaic modules, biogas digesters, small wind turbines; storage devices, flexible loads, power conditioners, and the management, operation and control equipment, interconnected and operated by electric and communication interfaces to satisfy the power necessities of a specific local community [15] [16].

### C. ICT and Enterprise Architecture

IT-architecture is the art and science of structuring and organizing information and systems. When there are many players having limited authority, different requirements, a variety of systems, different communication means, etc., it is not possible to *engineer* the situation because it is too complex, instead it is necessary to architect. Architecting focuses on ill-structured problems and on the need to create a shared view on what the future landscape should look like [17]. It is not possible to obtain an optimum, but it is possible to use heuristics to improve the landscape.

Enterprise Architecture (EA) extends the Information and Communication Technologies (ICT) architecture to the business process to guide design decisions by defining the system from its composition, dependencies among its elements, and the complexity involved. This EA is considered a master plan and a SoS. A good architecture contains both descriptive and prescriptive elements. An Enterprise Architecture cycle and meta-framework is presented in [17].

The analysis of ICT and enterprise architecture is useful in the context of microgrids because they are complex systems that have multiple, interconnected elements and different interoperability layers; thus finding a unique optimal solution of the complete system would not be suitable. The application of these insights is reflected with the specific design of the MRM.

## III. MICROGRID REFERENCE METHODOLOGY

The proposed Microgrid Reference Methodology (MRM) is expanded from [18] and modified for problems concerning the cooperation between utilities and industrial customers. The MRM shows an initial integration of the areas of knowledge reviewed in the previous section.

Initially, the main phases in the microgrid system life cycle are defined. Because a microgrid can be considered as a System of Systems (SoS), it can be analyzed using the methods and approaches defined for modeling a SoS [19]. However, the focus of this research relies on the factors that allow to successful microgrid implementation for the mutual benefits of its actors. The purpose is not just modeling the system as it is, but also considering all its phases, from planning to operation. A microgrid system will change its behavior according to different social, technological, economical and regulatory factors constantly in flux with the market. For this reason, the life cycle shown in Fig. 1 includes aspects of the SE and SoS adapted to the microgrid context.

This life cycle is sequential but not unidirectional. The iterative nature of the system enables continuous improvement through feedback loops after obtaining preliminary results and consulting back with the stakeholders. In addition, it is important to have in mind that verification and validation processes are important in each stage of the cycle to improve the correctness and usefulness of the model.

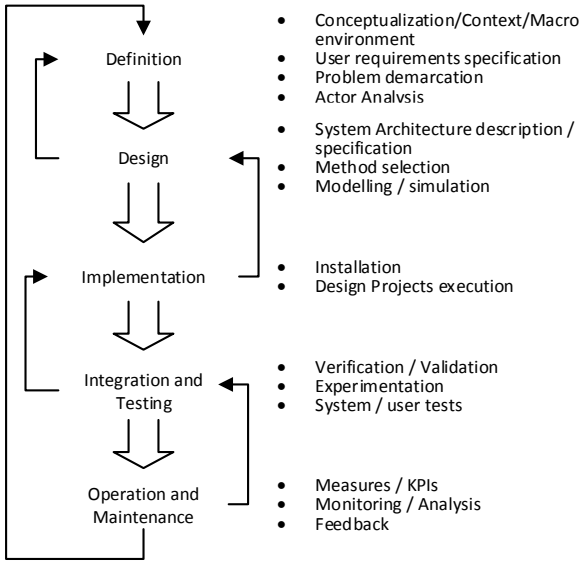

Fig. 1. Microgrid Reference Methodology Life Cycle Phases.

The Definition and Design phases constitute the innovation introduced in the MRM, and they focus on the planning and modeling of microgrids. The implementation, integration and testing, operation and maintenance phases are beyond the scope of this work because they are executed in real implementations once the design phase has been completely verified and validated by the stakeholders and the problem owner. In addition, no further observations are necessary to be made about these stages because the methods and guidelines are generic and very well documented by the Systems Engineering and Project Management bodies of knowledge.

*A. Definition*

The definition phase aims to understand the problem situation and the microgrid context for analysis. It is necessary to understand all the dimensions, the environment, stakeholders, interrelationships, interests, goals, etc. Hence, the collection of relevant information is crucial to accurately define this context.

Fig. 2 is a flow chart of the definition phase process. The first step is to collect information related to the dimensions and levels of the microgrid system and organize it using a system lexicon to maintain a common language within the microgrid project actors. It is used a table similar to the Resources, Operations, Policy and Economics (ROPE) table proposed in [20], with dimensions that are modified to include some factors of the Political, Economic, Social, and Technological (PEST) analysis. PEST is a tool used in strategic planning to identify the microenvironment and external forces of an organization focusing on the four previous environments.

In Table I is showed just two levels of The ROPE table to show a comparison between the Delta (δ) - Business and the alpha (α) – Technical level.

Fig. 3 shows the four levels of this Microgrid Reference Model. The α level has the largest number of elements, including equipment, devices, feeders, data, software, personnel, etc. These elements represent the initial considerations of the actors interested in developing microgrid projects. Traditionally, these elements have taken most of the time and attention when planning a new microgrid project. In addition, several efforts have been made in R&D by different developers and universities to obtain better, cheaper, and more efficient technology.

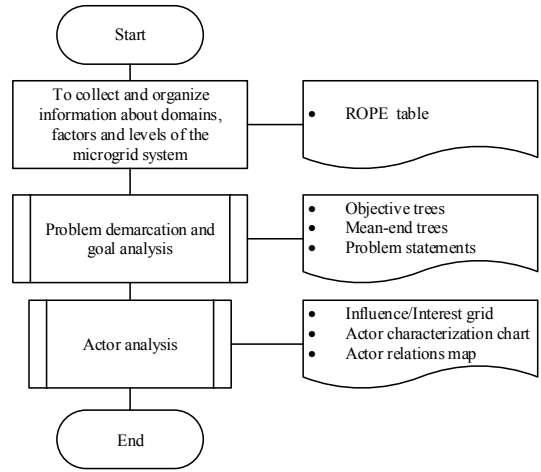

Fig. 2. Definition phase flow chart.

The next two levels have fewer elements, but the resources, operations, and factors involved are at a higher level. The focus of these layers is the efficiency of different systems inside the microgrid infrastructure and its operations. These levels are mostly considered when planning projects because they govern the individual interests of the microgrid owner and its senior management.

Traditionally, the first three levels have been analyzed to improve mostly technical aspects of the microgrid and some individual business goals of the microgrid owner, but not the interests of all actors at the same time. On the other hand, the δ level has the smallest number of elements, mostly organizations in the market and no technical components, so it has not been formally analyzed as an additional level in microgrid project design. However, the actors involved in this layer have very powerful interests, influence, and decision-making capabilities. The complexity of this level is even higher than the others, because there are socio-technical factors and variables. Hence, the lack of a complete understanding of the problem results in decisions that are not fully informed, and sometimes the execution of the microgrid project is not carried out once the technical concerns have been addressed.

The current MRM focuses on the δ level with a special interest in the cooperation between utilities and customers; however, the Definition phase specified in this chapter would be very useful for any problem involving any layer of the microgrid system. It is crucial to collect relevant information on the specific context of each microgrid project, identify direct stakeholders and actors, and determine their initial interests and concerns related to technical, regulatory, financial and business issues.

The next step in the Definition phase is the Problem Demarcation and Goal Analysis as shown in Fig. 4, which helps to identify higher-class and lower-class goals of the actors, the means to achieve them, and the undesired effects that must be controlled. This process is defined based on the methods and concepts proposed in [21] and [22]. By asking to the stakeholders' questions such as why is this important?, what is this?, and how can this be achieved?; hierarchical goal trees and means-end trees are created for each actor obtaining clear problem statements that considers goals and dilemmas.

Table I
ROPE table for microgrid systems (2-level comparison).

|  | α -Technical | δ – Business |
|---|---|---|
| Resources | Microgrid elements and devices (e.g. batteries, solar panels, AC/DC converters, loads, etc.) | Resources in the Microgrid/Energy Market |
| Stakeholders | Technical personnel (e.g. installers, engineers, etc.) | Organizations in the Microgrid market (e.g. utility companies, industrial customers, microgrid developers, regulators, etc.) |
| Operations | Operating a single technical resource (e.g. PV energy conversion, relay tripping, etc.) | Operations of Energy sector (e.g. implementing of incentives, defining of rate/tariff structures, billing & management, commercialization, etc.) |
| Policies | Policies relating to technical resources (e.g. standards, certifications, electric specifications, etc.) | Policies relating to the Energy Market (e.g. electric service tariffs, rules and regulations, Federal Energy Regulatory Commission acts, etc.) |
| Factors | Financial/ Technical/ Market / Social Concerns relating to single resources (e.g. efficiency in PV energy conversion, costs of fuel for a micro generator, etc.) | Financial/ Technical/ Market / Social Concerns relating to the energy sector (e.g. Profits, ROI, environmental impact, social welfare, etc.) |

The last step in the Definition phase is to perform a detailed actor analysis. Actors can be individuals or organizations, so an actor analysis helps to understand who is involved in the problem, who can influence the achievement of objectives, and their respective concerns and issues. A systematic process to identify the actors involved in the system was developed using the approach in [23] and [22]. The actor analysis results in a clear specification of the actors, their importance, and relationships in the system. An actor analysis helps to identify who has interests and who can influence the microgrid project.

In addition, the identification of the actors involved is useful to distinguish and understand the allies and the opponents. It is also useful in understanding the levels of power, resources, and interdependencies of different actors.

After defining the context, problem demarcation, and actor identification, the next phase is to develop a visual representation of the system and frame its dynamics using the design process.

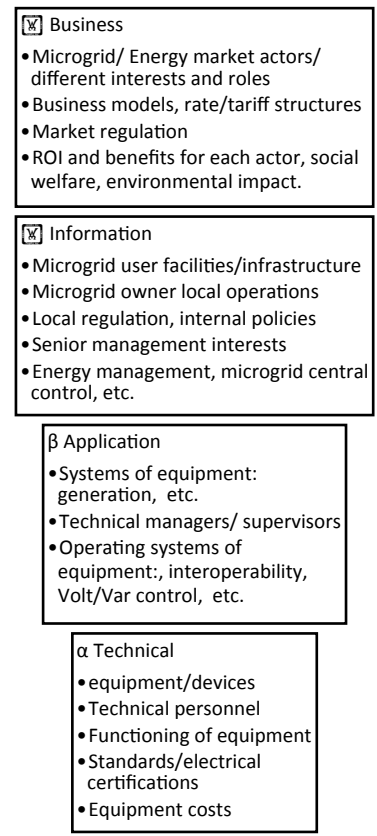

Fig. 3. Levels in the Microgrid Reference Methodology.

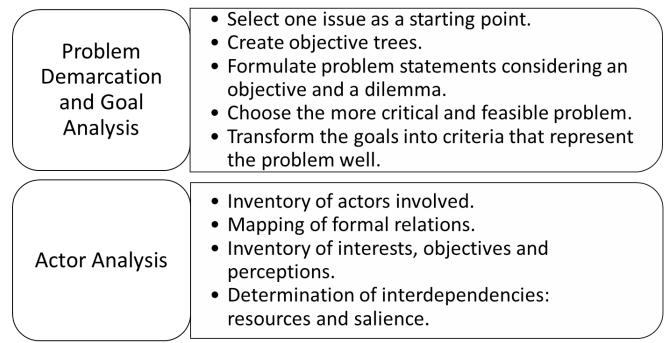

Fig. 4. Problem Demarcation, goal analysis and actor analysis.

## B. Design

The design phase aims to frame and architect the microgrid system by describing its actors' objectives, criteria, factors, interrelations, and the links between them. In addition, this phase aims to use the abstraction of the system, to define ways to measure its performance in different scenarios, and to provide alternatives for the stakeholders and decision makers. Furthermore, considering the cooperation between utilities and customers in this research, a systems dynamics approach based on the problem solving and decision making processes developed in [22] is used.

It is worth noting that the methods used in the design stage will vary depending on the problem and levels of interest. For example, if the problem lies at the α level, it might be better to characterize every element and device as an agent and use Agent Based Modeling (ABM). However, if the focus is the operations and processes of the microgrid, such as energy production and consumption over time, a discrete event approach might be more suitable. A flow chart of the general Design phase process is shown in Fig. 5.

After characterizing the problem statement, goals, and actors, the next step is to determine the factors that may influence the criteria and establish a causal relationship to finally create a problem diagram that represents the dynamics of the system and includes the objectives, criteria, factors and causal relations.

The next step is to consider external factors that cannot be influenced or controlled by the electric utility or the customers. These external factors generate different scenarios for evaluation based on the uncertainty degree and impact in the system.

Finally, the impact of different alternatives is evaluated in the accomplishment of the objectives for both, the electric utility and the customer. The best way to do this is using scorecards for each scenario. Scorecards are tables that show the effects of all alternatives on all criteria. The criteria are placed in the first column and the alternatives in the first row.

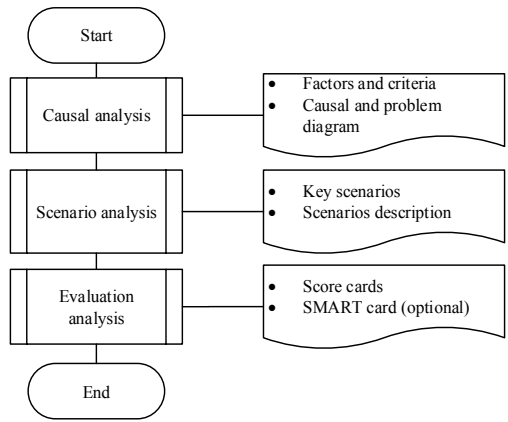

Fig. 5. Design phase flow chart.

## C. MRM application

In order to validate and reinforce the development of the Definition and Design phases of the MRM, information from primary and secondary sources was gathered and processed. Literature review, questionnaires, and interviews with representatives from some utility companies and industrial customers were utilized to determine the interests, concerns and objectives of each actor.

The insights obtained revealed that the main goals of customers in microgrid projects are: energy security in the event of a power outage, energy savings using renewable energies and batteries, and higher power quality to reduce spikes. On the other hand, utilities are concerned about limited existing incentives to provide any sort of ability to the customers to island themselves; costs to provide support power to the microgrid during peak hours; the necessity for changes in rate tariffs as consequence of reductions in power sales caused by customers using DER and microgrids; how to evaluate the environmental value of implementing microgrids; what to do in low electricity price environments such as the Midwest; the lack of U.S. standardized regulation of microgrids and disincentives for utilities to permit them; different prices and regulations in each state, new business models needed, etc.

The application of the problem demarcation uncovered the next problem statements:

• For electric utility company: how to obtain efficiency in generation usage without a reduction in profits?

• For industrial customer: how to reduce the impact of energy issues without incurring in high investments?

To visualize the interests and power of influence of each customer in a microgrid system, an influence/interest grid is depicted in Fig. 6. It shows that the players -those who have high power and high interest- are the utility company and the customer; however, it is important to consider the other actors' positions, because they might alter the decision of the microgrid project at some point during the project's life cycle.

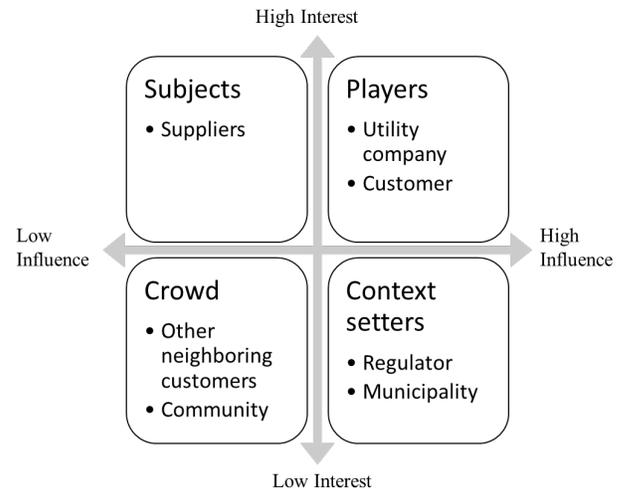

Fig. 6. Influence/Interest table of actors in Microgrid system.

The map of relations between the system's actors is shown in Fig. 7. For example, the regulators relate unidirectionally to the utility company and the customer because after the regulations are set, the regulators enforce them uniformly. In the case of the suppliers, they have bidirectional interactions with the utilities and customers in a business relationship, but their influence in the decision of implementing a new microgrid system is low compared with other actors. This map can apply for a current generic microgrid system; however, the interactions might change in specific cases and scenarios.

The causal and problem diagram for microgrid systems is depicted in Fig. 8 to show the factors, criteria and relationships uncovered as part of the application of the MRM for cooperation between electric utilities and customers. For example, an increase in the customer generation capacity decreases the customer electricity consumption form the grid. Those savings in electricity purchased might lead to improve the return on investments (ROI) and reduce the payback period of the customer. In this case, this event will contribute to meet the dilemma of the customer; however, the increase in the customer generation capacity also will increase the costs of new infrastructure for the customer reducing the ROI. In addition, an increase in the customer electricity consumption from the grid will reduce the revenue per kW produced by the electric utility as lesser efficient power generation is incrementally added, and this is an undesirable effect for the utility company.

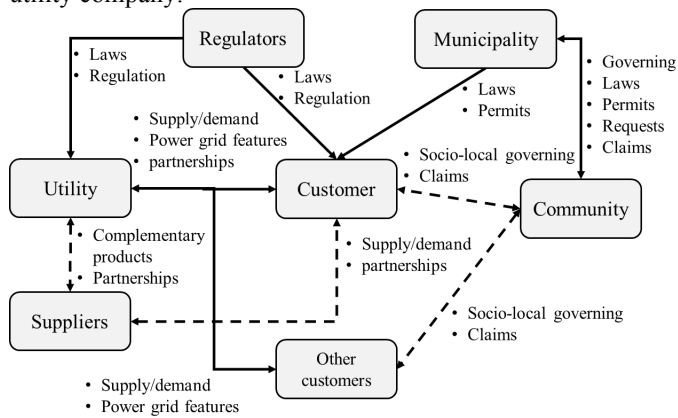

Fig. 7. Map of relations between actors in a Microgrid system.

Furthermore, there are more changes in other factors caused by the increase in customer generation capacity. For instance, the increase in costs of new infrastructure of the customer implies an improvement in generators, storage capacity and Management, Operations, and Control Systems (MOCS) which might lead to an increase in power quality, power reliability and resilience leading to a decrease in the cost of electric service issues which is the actual goal of the customer.

A change in a factor can influence several other factors leading to desired and undesired effects in meeting the actors' goals. Therefore, the possibility of interrelate those influences and to measure the impact helps to evaluate different alternatives and to take decisions based in quantitative variables.

Different alternatives can be proposed by the electric utility and the customer in a microgrid implementation, some of them thought by themselves and some analyzing the problem diagram trying to influence some desired factors. For instances, some alternatives can be: built a new infrastructure financed by the customer, built a new infrastructure financed by the utility, share investments and co-own infrastructure, develop new products and services for customers by the electric utility, etc.

Evaluating the alternative of building the new infrastructure financed by the customers for their own purposes results in a very low ROI, low cost of electric service issues, moderate load factor and very low revenue per kW produced. This is the least favorable to the electric utility and the customer because it does not solve the dilemma. The electric utility company would reduce considerably its incomes during the lifetime of the project, which usually is around 20 - 25 years with no direct benefit from not owning the microgrid. In addition, this option requires the customer to incur a large initial capital investment, which is sometimes higher than the savings in energy purchased from the grid. The lack of understanding of the mutual cooperation captured by the business level in the MRM causes several microgrid projects to be rejected. For this reason, when there is social welfare involved, most of the microgrid projects implemented until now have been financed by governmental incentives without the cooperation of the electric utilities.

After analyzing the impact of the alternatives for each criterion and the causal effects between the factors shown in the problem diagram, a mix of alternatives might be more beneficial. For example, combining the alternatives related to the new infrastructure financed by the utility and developing new products and services for customer it is possible to influence more factors and obtain desired effects. This will lead to a situation in which the customer's grid energy consumption would be reduced moderately. However, the customer does not need to worry about investments and initial capitals. In addition, the quality of energy and resilience of the network will be improved considerably, thereby achieving the customer's goal. The utility company in turn can benefit from the provision of ancillary services for this customer and neighboring customers to increase its utility revenues. Finally, a reduction in the peak hour demand will meet the electric utility company's goal of efficiency in generation usage.

The tools developed with this MRM allow us uncover other factors and alternatives that might be controlled by other actors that are not active players in the current scenario. For example, the level of DER technology development, regulatory restrictions and Standardization constitute external factors because neither the electric utility nor the customer can control them. Consequently they lead to different analysis scenarios. However, utilizing the outcomes of the actor analysis, shows that there are some factors, especially those concerning to regulation, that can be controlled and changed by other actors in the system. Actors such as equipment producers, universities, standardization organizations, and regulators directly impact those factors and create a scenario more attractive for investment and research in alternative energy generation with rules favorable to implement and commercialize innovative products and solutions.

These external factors can be influenced by alternatives that might not be feasible at the moment or in the current market conditions. These alternatives might involve important changes in the market and regulatory arena that would incentivize the implementation of microgrid projects and DER. These changes in regulation should not necessarily be solely based on the criteria and impact on the private objectives of electric utilities and customers, but also on social welfare. Energy efficiency, energy security, and environmentally friendly energy sources produce better quality electric service, and reduced environmental damage which are beneficial to society in general. Furthermore, the effect of these alternatives may benefit as well suppliers, communities, neighboring customers, and new actors such as information collection and processing to understand and describe the microgrid system, its context, and its actors. Unlike an unstructured and empirical negotiation process between these two key actors, as is typically used; this methodology considers the microgrid market as an additional level of the system. Hence, a systematic approach is used to identify key actors, interests, goals, and interrelations at this level. Important here as well is that this MRM offers the possibility of evaluating different alternatives in achieving the objectives of both electric utility and customer.

The MRM presented is perfectible with time. Continued research, case study applications, validations, and testbeds will enhance this methodology. The MRM creates a general architecture model for analysis, description, and microgrid operators and distribution supplier operations.

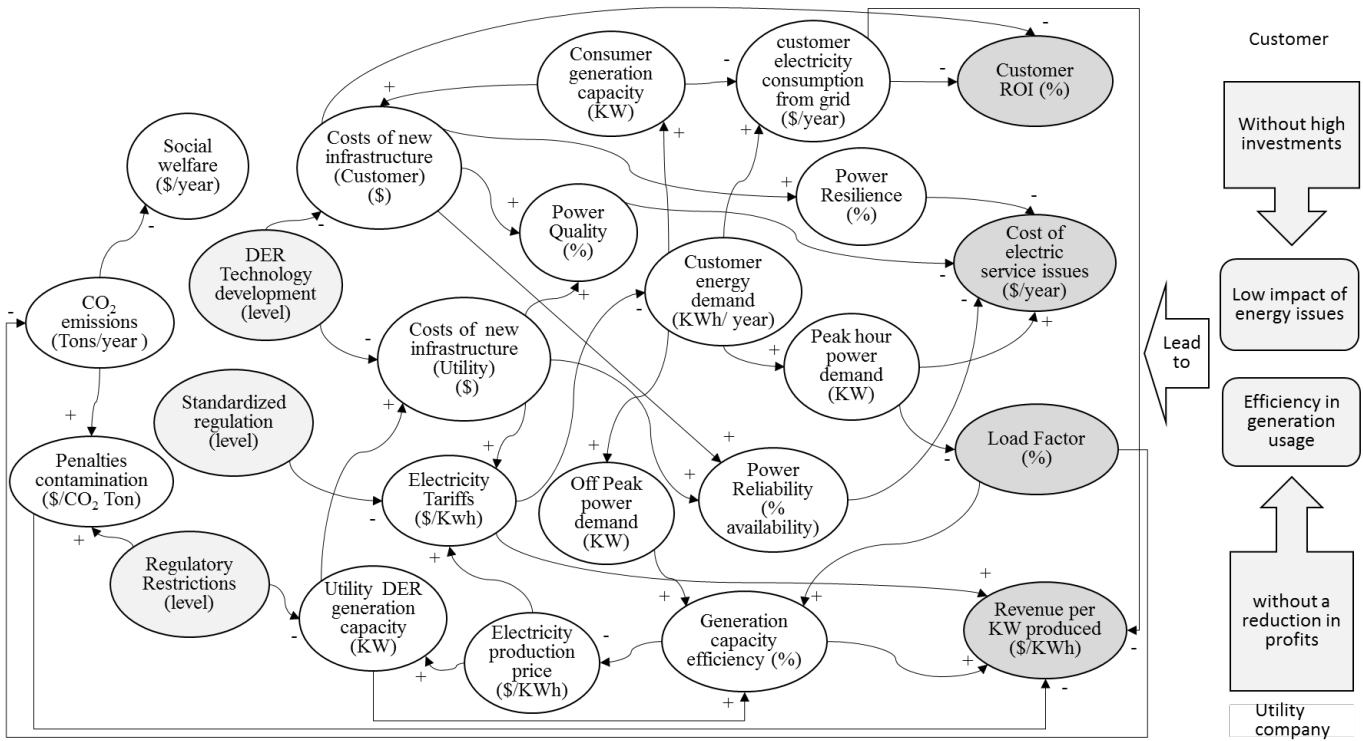

Fig. 8. Causal and problem diagram of the microgrid system.

It is important to mention that the application of the MRM requires the active participation of both electric utility and customer decision-makers. The process must be performed iteratively to achieve incremental improvement in the model. The first steps and stages of the MRM produce preliminary results, but the actors' feedback add depth and breadth to the analysis. Likewise, when the work proceeds to the next stage, it may be necessary to return to the previous steps to improve the description and understanding of the system.

IV. CONCLUSIONS

The Microgrid Reference Methodology proposed in this paper provides the framework for a systematic analysis of the interactions between electrical utility companies and customers for their mutual benefit. This MRM guides standardization of operation and information flows through the various dimensions, zones, and levels of the Microgrid System of Systems.

An important area of future study is microgrid modeling and and simulation. Currently there are different tools for general system simulation and specialized tools for power systems that focus more on the three lower levels featured in or rejecting a microgrid project. Therefore, there is need for integral computer tools that simulate the microgrid as a complete system, and represent its complexity, behavior, and state variables for each level under different conditions. This will reduce the time necessary to design and analyze new microgrid projects, and facilitate the decision making process in order to reach better agreements between the different actors.

Fig. 3. Existing commercial tools for power systems do not consider the microgrid market as an additional level, as we incorporated into this research. Hence, this model, based on more complete information about actors, goals, factors and the interrelationships in the all levels of a microgrid system will produce better results. In addition, this will identify hidden benefits and costs that are not determined with current tools and sometimes led to incorrect decisions about justifying